\documentclass[review,number,sort&compress]{elsarticle}

\usepackage{amssymb}
\usepackage[ngerman,english]{babel}
\usepackage{lscape}
\usepackage{rotating}
\usepackage{units}
\usepackage{psfig, epsfig}
\usepackage{subfigure}

\def \astri {ASTRI SST-2M}

\def \deg {$^{\circ}$}

 \usepackage{lineno}

\journal{Nuclear Physics A}

\begin{document}

\begin{frontmatter}


 
 \title{Volcanoes muon imaging using Cherenkov telescopes}

\author[label1]{O. Catalano}
\author[label1]{M. Del Santo}
\ead{melania@ifc.inaf.it}
\author[label1]{T. Mineo}
\author[label1]{G. Cusumano}
\author[label1]{M. C. Maccarone}
\author[label2]{G. Pareschi}
 
\address[label1]{INAF, Istituto di Astrofisica Spaziale e Fisica cosmica di Palermo, via U. La Malfa 153, I-90146 Palermo, Italy}
\address[label2]{INAF, Osservatorio Astronomico di Brera, Via E. Bianchi 46, I-23807, Merate, Italy}



\begin{abstract}
A detailed understanding of a volcano inner structure is one of the key-points for the volcanic hazards evaluation.
To this aim, in the last decade, geophysical radiography techniques using cosmic muon particles have been proposed.
By measuring the differential attenuation of the muon flux as a function of the amount of rock crossed along different directions,
it is possible to determine the density distribution of the interior of a volcano.
Up to now, a number of experiments have been based on the detection of the muon tracks crossing hodoscopes, 
made up of scintillators or nuclear emulsion planes. 

Using telescopes based on the atmospheric Cherenkov imaging technique,
we propose a new approach to study the interior of volcanoes detecting of the Cherenkov light produced by relativistic cosmic-ray muons that survive
after crossing the volcano. 
The Cherenkov light produced along the muon path is imaged as a typical annular pattern containing  
all the essential information to reconstruct particle direction and energy. 
Our new approach offers the advantage of a negligible background and an improved spatial resolution.

To test the feasibility of our new method, we have carried out simulations with a toy-model based on the
geometrical parameters of  \astri, i.e. the imaging atmospheric Cherenkov telescope currently under installation onto the Etna volcano.
Comparing the results of our simulations with previous  experiments based on particle detectors, we gain at least a factor of 10 in sensitivity.
 The result of this study shows that we resolve an empty cylinder with a radius of about 100 m located
inside a volcano in less than 4 days, 
 which implies a limit on the magma velocity of 5 m/h.

\end{abstract}

\begin{keyword}


muon radiography \sep  volcano structure \sep  Imaging Atmospheric Cherenkov Telescope \sep  Silicon Photomultiplier 
\end{keyword}

\end{frontmatter}


\section{Introduction}
\label{intro}

Cosmic ray muons are created when high energy primary cosmic rays interact with the Earth' s atmosphere  (see \cite{nagamine03} for a recent review). 
Muon imaging for non-destructive studies of gigantic objects arose soon after their discovery.
Muon radiography was first proposed to determine the thickness of a snow horizontal tunnel on a mountain in Australia \cite{george55}.
Then, in archeology, it was adopted to investigate the interior of the Egyptian pyramid of Chephren at Giza in order
to find a hidden chamber \cite{alvarez70}. 
Very recently the muon tomography technique has been used to inspect the content of traveling cargo containers \cite{larocca15}.
In 2007, H. Tanaka and collaborators from the University of Tokyo were the first to apply this technique to volcanoes \cite{tanaka07}. 

Volcanic eruptions occur when magma from the inner of the Earth comes out on the surface through the main conduit or conduits 
connecting the underground magma reservoir with the erupting crater. 
The eruptions comes out from the volcano's mouth or from a number of mouths that open at different points. 
The duration of volcanic eruptions is variable: they may last a few hours or even decades. 
Although the interpretation of geophysical signals emitted from volcanic conduits such as micro-seismicity, surface deformation and
gas and thermal emissions are used for eruption forecasts, only little direct information is available on the conditions inside and near the conduit of an active volcano. 
Improvements in measuring the size of the conduits could help in the interpretation of premonitory marks and in the risk reduction. 

The density distribution of the interior of a volcano has been determined by measuring the differential attenuation of the muon flux 
as a function of the amount of rock crossed along different directions. 
So far,   measurements in this context have been based on the detection of the muon tracks crossing hodoscopes, 
made up of scintillators or nuclear emulsion planes. 
However, this technique requires several detection layers and a sufficiently high timing resolution to reduce the level of fake coincidences due to the unavoidable charged particles background.\\

We present for the first time the feasibility study of using an imaging Cherenkov telescope to carry out 
the muon radiography of a volcano.
We aim to apply this technique to the Etna volcano 
using the \astri\ telescope currently under installation.
The advantage of using Cherenkov telescopes for muon radiography is 
due to their imaging capability which results in
negligible background and improved spatial resolution
compared the traditional particle detectors.


\begin{figure*}[t!]
\begin{center}
\centering
\psfig{figure=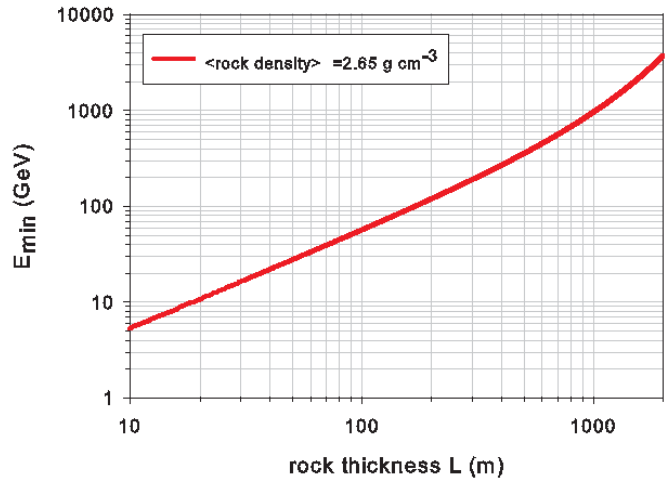,height=7cm}
\end{center}
\caption{
Energy threshold for a muon to cross a rock with standard density (2.65 g cm$^{-3}$) at different thicknesses. The curve results from Eq. \ref{eq:e_mu_min}. 
}
\label{fig:emin1}
\end{figure*}

\section{Muon radiography: observational approach}
\label{muons_ima}

Muon imaging allows us to determine the density variations in the inner structure of a volcano by measuring
the differential attenuation of the muon flux. 

\subsection{Principles of muon imaging}
The flux of atmospheric muons incoming the volcano at a given arrival direction can be determined using Monte Carlo simulation codes \cite{heck98} 
or fitting data of experimental results obtained looking at the open sky at the same angle \cite{bugaev98}. 

Any muon flux variation translates in a difference in the opacity ($X$) which is defined as:
\begin{equation}
X(L) \equiv \int_{L} \rho(\xi) \ {\rm d}\xi \ (\rm{g \ cm^{-2}})
\label{eq:flux}
\end{equation}
where $\rho$ is the rock density and $\xi$ is the spatial coordinate measured along the trajectory $L$ of the muon crossing the rock.


 The integrated cosmic muon flux after the volcano has been crossed (as a function of the opacity $X$ and of the zenith angle $\theta$) is defined as: 
\begin{equation}
I(X, \theta)=\int_{E_{\rm min}}^\infty  J(E, \theta) \ {\rm d}E\  ({\rm cm}^{-2} \ {\rm sr}^{-1} \ {\rm day}^{-1}).
\label{eq:flux}
\end{equation}
\noindent
where  $J(E, \theta) \equiv dN(E,\theta)/dE$  (cm$^{-2}$  sr$^{-1}$  day$^{-1}$  GeV$^{-1}$) is
the differential flux of incident muons at a given angle.
The minimum muon energy ($E_{{\rm min}}$) required to cross a depth of opacity $X$  is calculated using 
the classical definition of the average muon energy loss: 

\begin{equation}
dE_{\mu}/dX = -a \ -bE_{\mu} 
\label{eq:de_mu}
\end{equation}
\noindent
with $a \approx 2$ MeV g$^{-1}$ cm$^{2}$ and $b \approx 4 \times 10^{-6}$ g$^{-1}$ cm$^{2}$ \cite{darijani14} 
for a rock with standard density (2.65 g cm$^{-3}$).

\begin{figure*}[t!]
\begin{center}
\centering
\psfig{figure=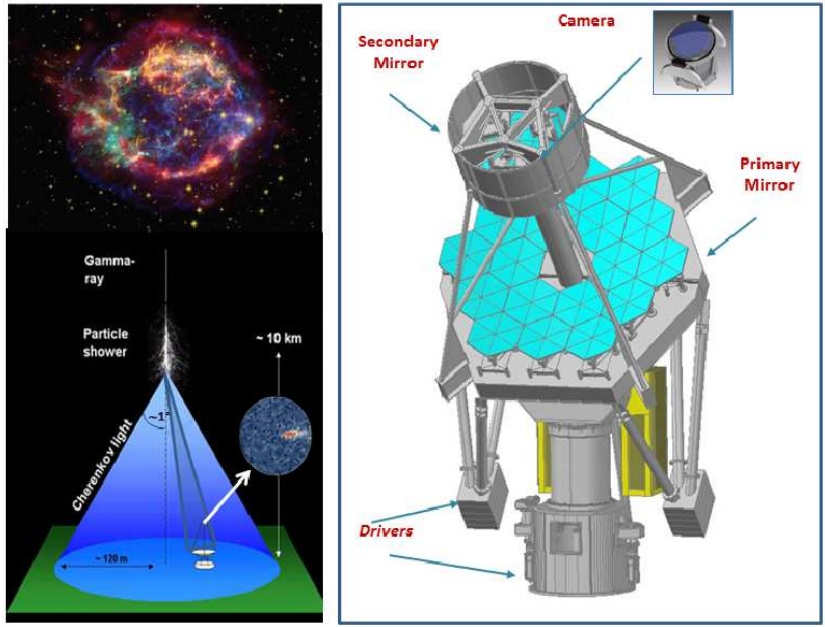,height=9cm}
\end{center}
\caption{
The Imaging Atmosphere Cherenkov Technique and the working principle of the \astri\ telescope. 
{\it Left}: the Cherenkov light detected by the telescope is produced by a gamma-ray interacting with the atmosphere.
{\it Right}: In the \astri\ telescope Cherenkov photons are reflected by the primary mirror onto secondary mirror which focusses them onto the camera.
}
\label{fig:working_astri}
\end{figure*}

\noindent
Using Eq. \ref{eq:de_mu}, the muon energy after propagation is:

\begin{equation}
E_{\mu} = (E^{0}_{\mu} + \epsilon) \ {\rm exp} (-bX) \ -\epsilon
\label{eq:e_mu_eps}
\end{equation}
\noindent
with $\epsilon=a/b$ and $E^{0}_{\mu}$ the energy of the incident muon.
Resolving Eq. \ref{eq:e_mu_eps} for $E_{\mu}=0$,  we obtain the minimum muon energy required to cross a depth with opacity $X$:

\begin{equation}
E_{{\rm min}} = \epsilon \ [{\rm exp} (+bX) -1].
\label{eq:e_mu_min}
\end{equation}

\noindent
Figure \ref{fig:emin1} shows the minimum muon energy corresponding to the propagation path in a standard density rock. 

The feasibility of the muon imaging to investigate the density distribution inside 
a target structure can be inferred through the relation suggested and exhaustively treated in \cite{lesparre10} : 

\begin{equation}
\Delta T \times \Gamma \times \frac{\Delta I^{2}(X_{0}, \delta X)}{I(X_{0})} > 1 
\label{eq:feasi}
\end{equation}
with $\Delta T$ the acquisition time, $\Delta I(X_{0}, \delta X) =  I(X_{0}+ \delta X) - I(X_{0})$, $X_{0}$ the fixed total opacity of the medium,
$\delta X$ the required resolution level, and
$\Gamma$  the detector acceptance defined as:
\begin{equation}
\Gamma = A \times 2\pi (1-\cos (\alpha/2))
\label{eq:accept}
\end{equation}
where $A$ is the detector geometrical area and $\alpha$ is the angular resolution.
Density and size of the volcano inner structure can be estimated with an angular resolution driven by the detector performance.
 
Eq. \ref{eq:feasi} establishes a useful relationship 
between the acquisition time necessary to collect a statistically significant number  of muons,  $\Delta T$,
and the integrated flux differences 
of  muons crossing different directions inside the target, $\Delta I(X_{0}, \delta X) =  I(X_{0}+ \delta X) - I(X_{0})$.
It is worth noting that Eq. \ref{eq:feasi} is applicable only in case of negligible background.

\begin{figure*}[t!]
\begin{center}
\centering
\psfig{figure=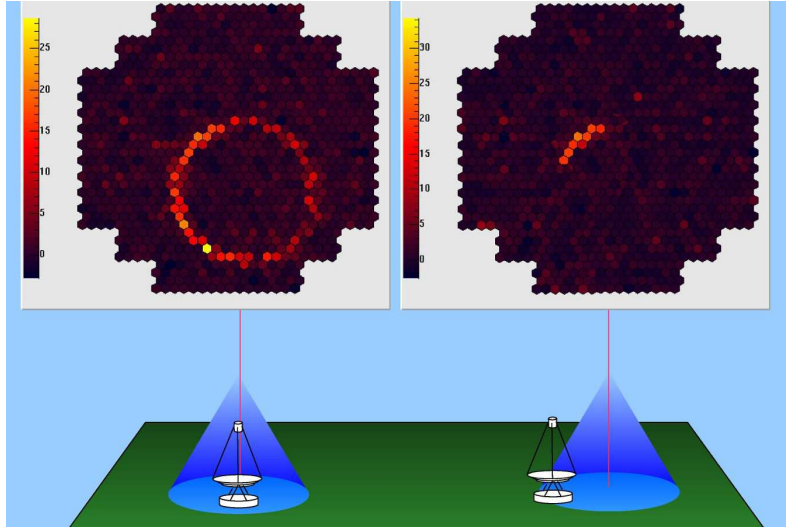,height=7cm}
\end{center}
\caption{Muon ring images with a Cherenkov telescope (H.E.S.S. in this case) when muon hits the mirror (left); when the impact point is outside the mirror
the image is an arc (right; the figure is published in \cite{voelk09}).
}
\label{fig:muhess}
\end{figure*}

\subsection{Experiments}
The instrumental approach  is currently based  on the detection of muons crossing hodoscopes made 
up of scintillator planes.
First results have been obtained by Tanaka and collaborators who carried out muon radiography of the top part of the Asama volcano in Japan  
and revealed a region with rock of low density on the bottom of the crater \cite{tanaka07, tanaka09}.
They definitely demonstrated the feasibility of the method to detect both spatial and temporal changes
 of density inside a volcano  \cite{nagamine95,tanaka01,tanaka03}.
Very recently, the first  muographic visualisation  of the dynamics of a magma column in an erupting volcano has been presented \cite{tanaka14}.
These authors performed a muon radiography of the Satsuma--Iwojima
volcano  showing that  while the eruption column was observed, the top of the magma column
reached a location of 60 m beneath the crater floor. 
Moreover, they proposed that the monitoring of the temporal variations in the gas volume
of the magma as well as its position in a conduit could be used to support eruption prediction.
Recently, a new telescope prototype has been proposed to study Mt. Vesuvius  \cite{ambrosi11}.
It is based on the use of bars of plastic scintillator with a triangular section whose scintillation light is collected by 
Wave-Length Shifting (WLS) optical fibers
and transported to Silicon photomultipliers \cite{anastasio13}. 

In the summer 2010, the first experiment of muon radiography at Mt Etna  
by using a detector employing  two scintillator planes was carried out \cite{carbone13}.
A marked difference between theoretical and observed attenuation of muons through
the crater was found. This discrepancy was likely due to the bias on the observed flux, arising from false
muon tracks. These are caused by muons arriving from isotropic directions and by
low-energy particles of ordinary air showers that, by chance, hit simultaneously the
two detector planes, leading to the detection of a false positive.

\section{Muons with Cherenkov telescopes}
\label{muon_tel}
Cherenkov light is emitted when charged particles, such as muons, travel through a dielectric medium with velocity ($v$) higher than the speed of light in that medium, 
i.e. $ v > c/n $ where $c$ is the speed of light in vacuum and $n$ the refraction index.  This implies an energy threshold 
for Cherekov production of $\sim 5$ GeV  in the atmosphere at sea level.
The light is emitted in a narrow cone with an angle cos $\Theta$= 1/$\beta$n ($\beta=$v/c)
around the direction  of motion of the particle, $\Theta$ $\approx 1.3^{\circ}$  in air. 
 The Cherenkov radiation relevant for detector application occurs in the visible and ultraviolet regions of the electromagnetic spectrum.

Cherenkov light produced by relativistic charged  particles in a shower induced by TeV photons  interacting with the Earth atmosphere
allows us to study very high-energy emission  (VHE; $>$0.1 TeV)
coming from sky-sources, i.e.  supernova remnants, pulsars, Active Galactic Nuclei, by using on-ground telescopes.
 Imaging Atmospheric Cherenkov Telescopes (IACT) have opened a window to the ground-based gamma-ray astronomy in the VHE
range and continuously improves in what concerns detection performance and sensitivity, 
using multiple telescopes in stereoscopy (VERITAS \cite{weekes02}, MAGIC \cite{tridon10}, H.E.S.S. \cite{benbow05}, currently in operation).
The next generation of ground-based VHE gamma-ray instrument is the Cherenkov Telescope Array (CTA; \cite{actis11}) 
which will provide a deep insight into the non-thermal high-energy Universe.
Basically, a IACT consists of an optical system formed by highly reflectivity mirror(s), 
that focuses the impinging Cherenkov radiation onto a multi-pixel camera equipped with a fast read-out electronics (Fig. \ref{fig:working_astri}).

\begin{figure*}[t!]
\begin{center}
\hbox{
\centering
\psfig{figure=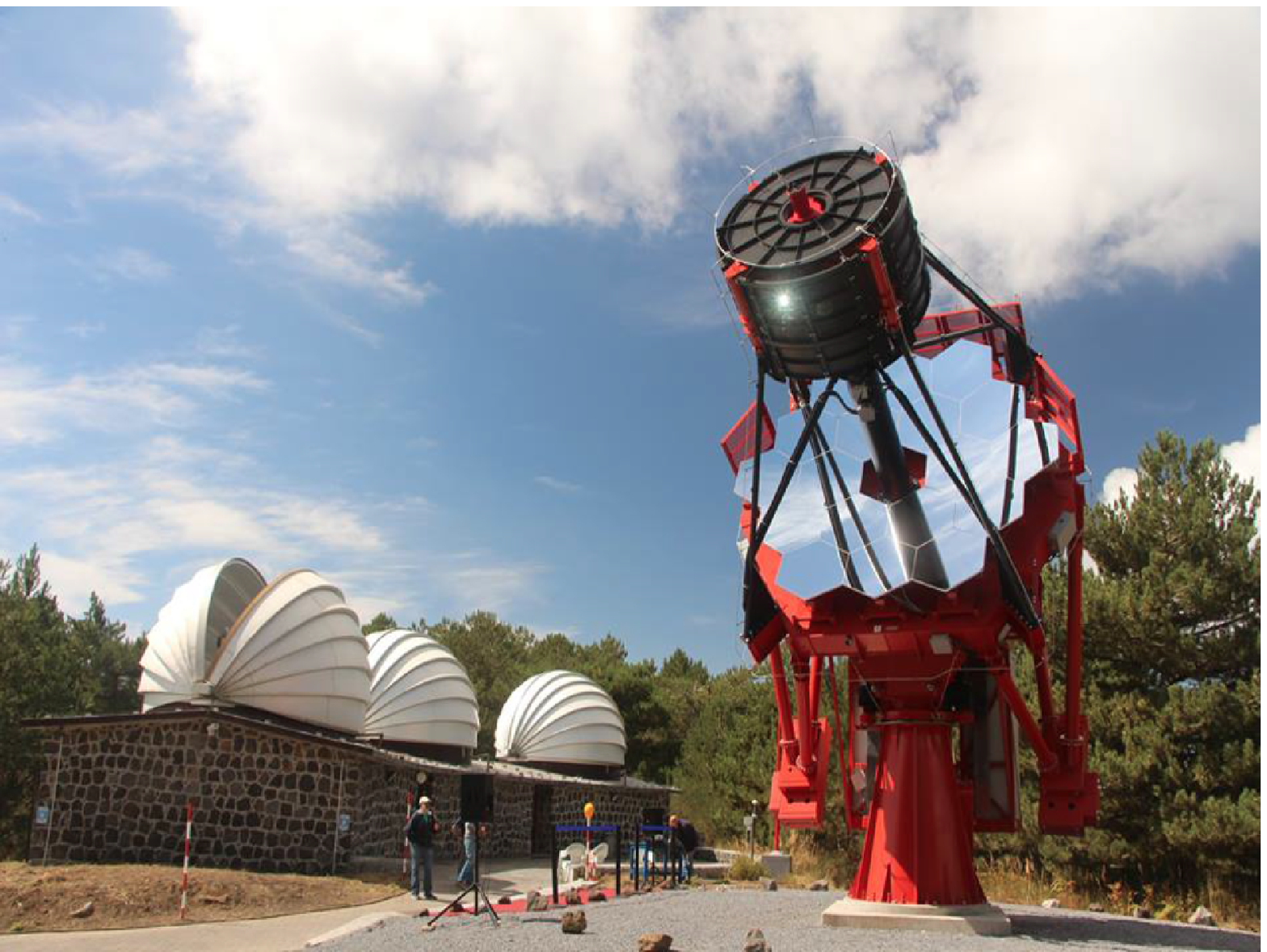,height=4cm,width=6cm}
\hspace{0.5cm}
\psfig{figure=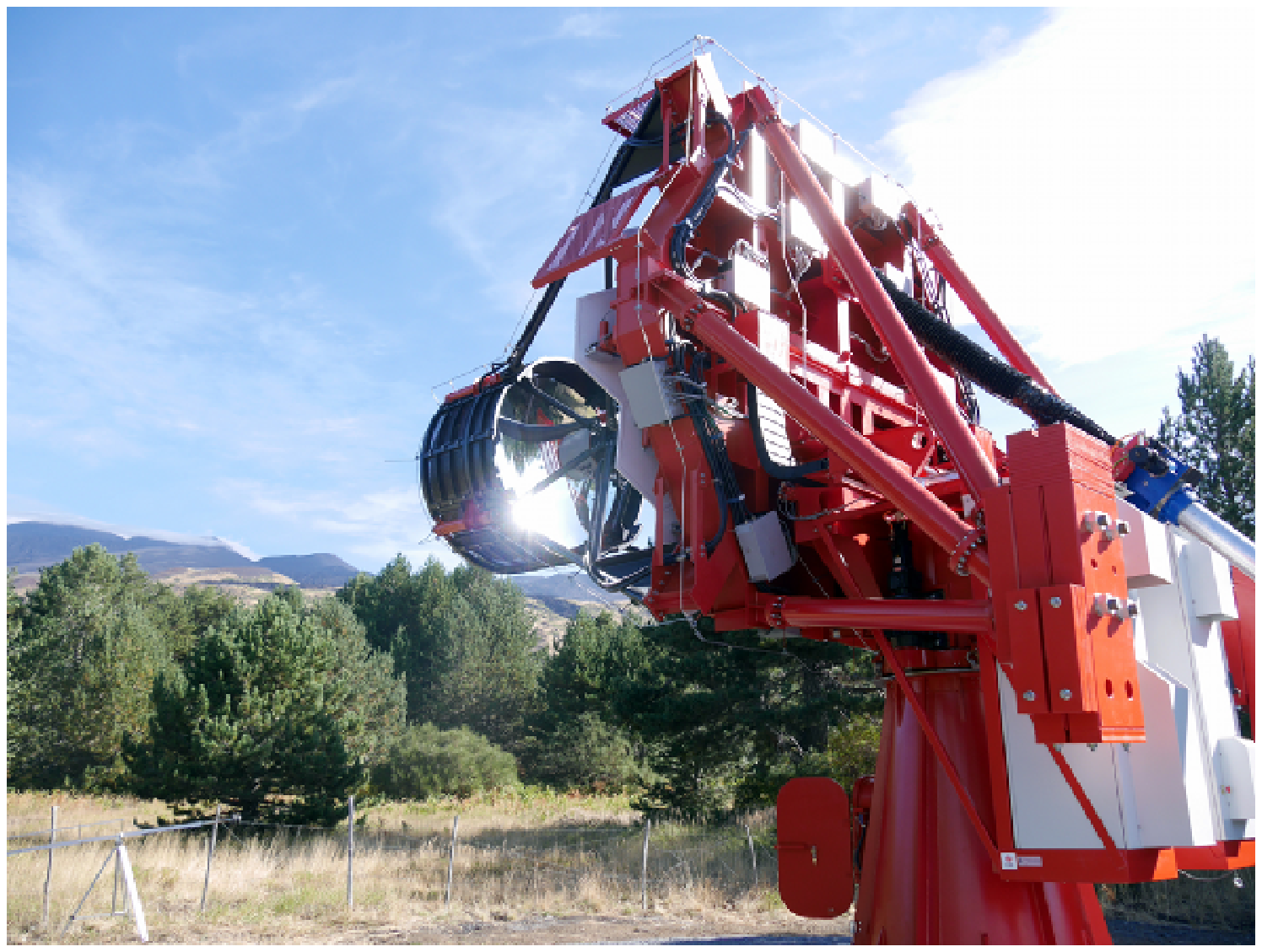,height=4cm,width=5cm}
}
\vspace{-1.0cm}
\end{center}
\caption
{\astri\ telescope installed at Serra La Nave observing station. 
In the right picture, over the trees, the Etna volcano is  visible.}
\label{fig:astri}
\end{figure*}


The unique characteristics of images produced by muons in IACTs make their detection
an useful tool  to calibrate their optical throughput \cite{vacanti94, rovero96, meyer05, bolz04}.
When a muon hits the mirror of an IACT,  the light emitted along the final part of its path
is imaged as a ring in the camera at the focal plane of the telescope (\cite{vacanti94}; see Fig. \ref{fig:muhess}). 
The amount of collected light mainly depends on the impact point of the muon on the telescope mirror, 
on the geometrical area of the entrance pupil of the telescope and on the camera efficiency. 

A relatively simple geometrical analysis of the ring allows us to reconstruct the muon physical parameters, i.e. its energy and arrival direction.
The position of the centre gives the muon arrival  direction respect to the telescope optics axis. 
The radius of the ring corresponds to the Cherenkov angle ($\Theta$) that depends on the energy of the muon ($E_{\mu}$)
according to the formula:
\begin{equation}
E_{\mu}= \frac{0.105} {\sqrt{1- (n \ \cos \  \Theta)^{-2} }} \  (\rm{GeV}). 
\label{eq:e_mu}
\end{equation}
\noindent
  It results that  the Cherenkov angle saturates with the increasing particle energy.
At an observation altitude of 1800 m, the Cherenkov angle
saturation occurs at an energy higher than $\sim$50 GeV.

\subsection{\astri}
ASTRI (Astrofisica con Specchi a Tecnologia Replicante Italiana) is a flagship project of the Italian Ministry of Education, 
University and Research, led by the Italian National Institute of Astrophysics (INAF) 
and developed in the framework of the ambitious CTA project.
The first task of the ASTRI project is the realisation of an end-to-end telescope prototype 
that has been proposed for the small-size class telescopes of the CTA. These will be devoted to the investigation of the sky in the energy range from a few TeV up to more than 100 TeV. 
The prototype, namely \astri, is characterised by a dual-mirror Schwarzschild-Couder optical design \cite{vercellone13}, 
adopted for the first time on a Cherenkov telescope and by an innovative modular camera at the focal plane
managed by a very fast read-out electronics \cite{catalano14}. 
The primary mirror of the telescope has a 4.2 m diameter and is composed by an array of hexagonal tiles, 
while the secondary optics is a monolithic 1.8 m diameter mirror \cite{canestrari14}. 
The camera consists of a matrix of Silicon Photomultiplier (SiPM) sensors covering a 9.6$^{\circ}$ full field of view (FOV) \cite{catalano13}, 
with a pixel solid angle  of $0.17 \times 0.17$ deg$^{2}$, and a Point Spread Function (PSF), defined as the 80\% of the light collected from a point like source, contained in one pixel.
 In Fig. \ref{fig:working_astri} ({\it right}), we show a sketch of the working principle of \astri.
During Autumn 2014, \astri\  has been inaugurated at the INAF "M.C. Fracastoro" observing station \cite{maccarone13, leto14} 
located in Serra La Nave (Mount Etna, Sicily;  Fig. \ref{fig:astri}).
At the time of this paper writing, the telescope structure, mirrors and control software are installed;
as soon as the camera is mounted, the telescope will be operative.
The main aim of this prototype is to characterise the performance  of such a telescope by observing the Crab nebula and two Markarian galaxies (Mrk421, Mrk501).

The location on Mt. Etna will offer us the opportunity to test through real data the use of a Cherenkov telescope
for the volcano muon radiography.
\astri\ will be able to detect muons by collecting in a single ring the light emitted along the last $\sim$100 m 
of its path towards the primary mirror.
According to our simulations, \astri\ will be able to reconstruct muons with a precision on the direction of about 0.14\deg,
if the muon energy is higher than 20 GeV and the muon hits the primary mirror up to an off-axis angle of 3.6\deg\ \cite{strazzeri13}.

\begin{figure*}[t!]
\centering
\includegraphics[width=12cm]{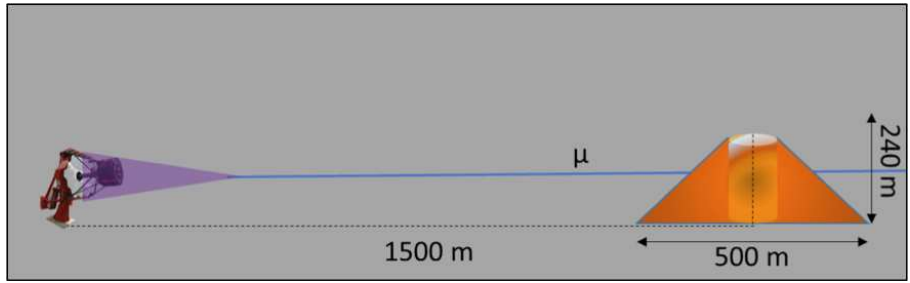}
\caption{Sketch of the telescope-volcano configuration used for our simulations. It is worth to note that Cherenkov light is produced along the whole muon path,
while we show only Cherenkov light useful for \astri\ which is that produced in the last $\sim$100 m.}
\label{fig:ast_sim}
\end{figure*}

\section{Muon radiography of Mt Etna with Cherenkov light: simulations}
\label{muon_rad}
To study the feasibility of detecting muons crossing the Mt Etna by Cherenkov light,
we carried out simulations with a toy model based on the \astri\ geometrical parameters  and simplified shapes for both the volcano and its conduit.  
In order to have the whole Mt Etna within the FOV of the telescope, 
the distance between telescope and volcano has been set at 1500 m (Fig. \ref{fig:ast_sim}).

With our set-up the background is due to muons hitting the primary mirror with an incidence angle within the FOV and not coming from the mountain. 
These events, mainly back-scattered from the ground,  would produce a ring and cannot be discriminated by the useful signal.
 Assuming the ground level measurement of upward directed atmospheric muons of 3 $\times 10^{-6}$ cm$^{-2}$ sr$^{-1}$ day$^{-1}$ (see Fig. 3.79 in \cite{grieder01}),
a rough estimation of the level of this muon background results in a rate of about 3$\times 10^{-3}$ "fake" events per observation night within the field of view of \astri.

\subsection{Experimental set-up}

The volcano geometry is represented by a simple cone of base 500 m and height
of 240 m roughly corresponding to the dimensions of the South-Est (SE) mouth of the Etna volcano.
Conduits of various dimensions are simulated as hollow cylinders
embedded into the  cone with  axis coincident with that of the cone. 
The cone is extended with an elevation angle between 90$^{\circ}$ (cone base) and 80.4$^{\circ}$ (cone summit) 
with the telescope optical axis pointing towards the half height of the cone axis (Fig. \ref{fig:ast_sim}). 
The cone has been segmented with bins of 0.17$^{\circ} \times 0.17^{\circ}$, corresponding to the  camera pixel angular size,
 although the actual \astri\ resolution for muon reconstruction direction is about 0.14$^{\circ}$ \cite{strazzeri13}.  
 According to this geometry, the projected spatial resolution referred to the axis cone is about 4.5 m.

\begin{figure*}[t!]
\begin{center}
\psfig{figure=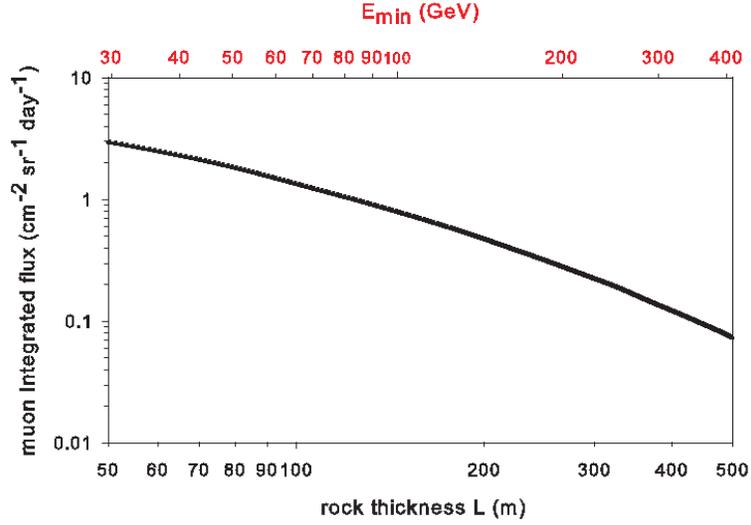,height=7cm}
\end{center}
\caption{
Muon integrated flux as a function of the standard rock thickness. The top axis gives the required minimum energy to cross the corresponding rock thickness
}
\label{fig:emin2}
\end{figure*}

We assume the integrated flux $I$($X, \theta$) computed by \cite{lesparre10}, 
for $\theta$=85\deg\ and for a density typical of the standard density rock,
expressed as a function of the crossed thickness $L$ (Fig. \ref{fig:emin2}).
The integrated flux is calculated starting from the Reyna-Bugaev model 
which allows to compute the differential flux of incident muon, $J(E, \theta)$, for all zenith angles and a wide energy range \cite{lesparre10}.
The feasibility condition is established by Eq. \ref{eq:feasi} which sets the minimum time 
to resolve the average density distribution inside the target.
In the \astri\ case the bin acceptance ($\Gamma$) is $\approx$1 cm$^{2}$ sr.

\begin{figure*}[t!]
\centering
\includegraphics[width=12cm]{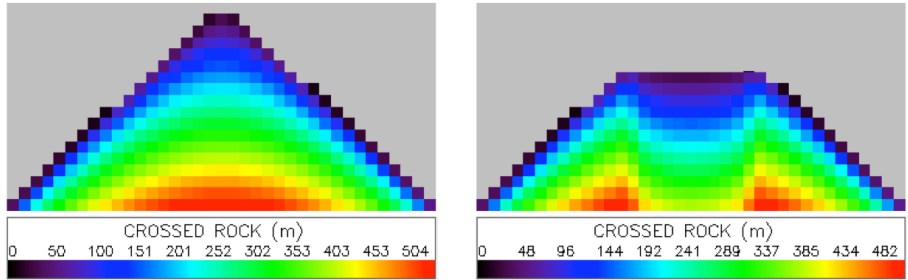}
\caption{Distribution of the crossed rock lengths in the full cone (left) and in the one modified adding a cavity of 144 m diameter in its interior (right). The grey area
on the background indicates the open sky.}
\label{fig:cone}
\end{figure*}

\subsection{Results}
\label{results}
The integrated flux of simulated muons crossing each bin has been recorded both for the full cone and for the hollow cylinder and Eq. \ref{eq:feasi} has been applied. 
We have then performed a grouping as 3$\times$3 bins,  corresponding to a new positioning resolution of 13.5 m 
and an acceptance of 9 cm$^{2}$ sr.
This allowed us to obtain a better statistics, to reach a telescope acceptance similar to that of a previous experiment \cite{carbone13},
and to have pixel size comparable with the muon angular deviation expected from multiple scattering in crossing the mountain \cite{nagamine03}.
Note that the scattering inside the volcano  depends on the energy of the muon. Therefore, muon angular reconstruction can be improved
by selecting events in energy.
From our simulations it results that the scattering effect is negligible for reconstructed energies higher than  30 GeV at the base of the volcano  and
 higher than 5 GeV on the top.

Figure \ref{fig:cone} shows the length distribution of the crossed rock for the full cone and for the cone with an empty cylinder inside (diameter D = 144 m). 
The reduction in the crossed rock paths along the cone axis  where the hollow cylinder is positioned is well evident (right panel).

\begin{figure}[!h] 
\noindent
\centerline{
  \hspace{-1.2truecm}
                  \epsfig{figure=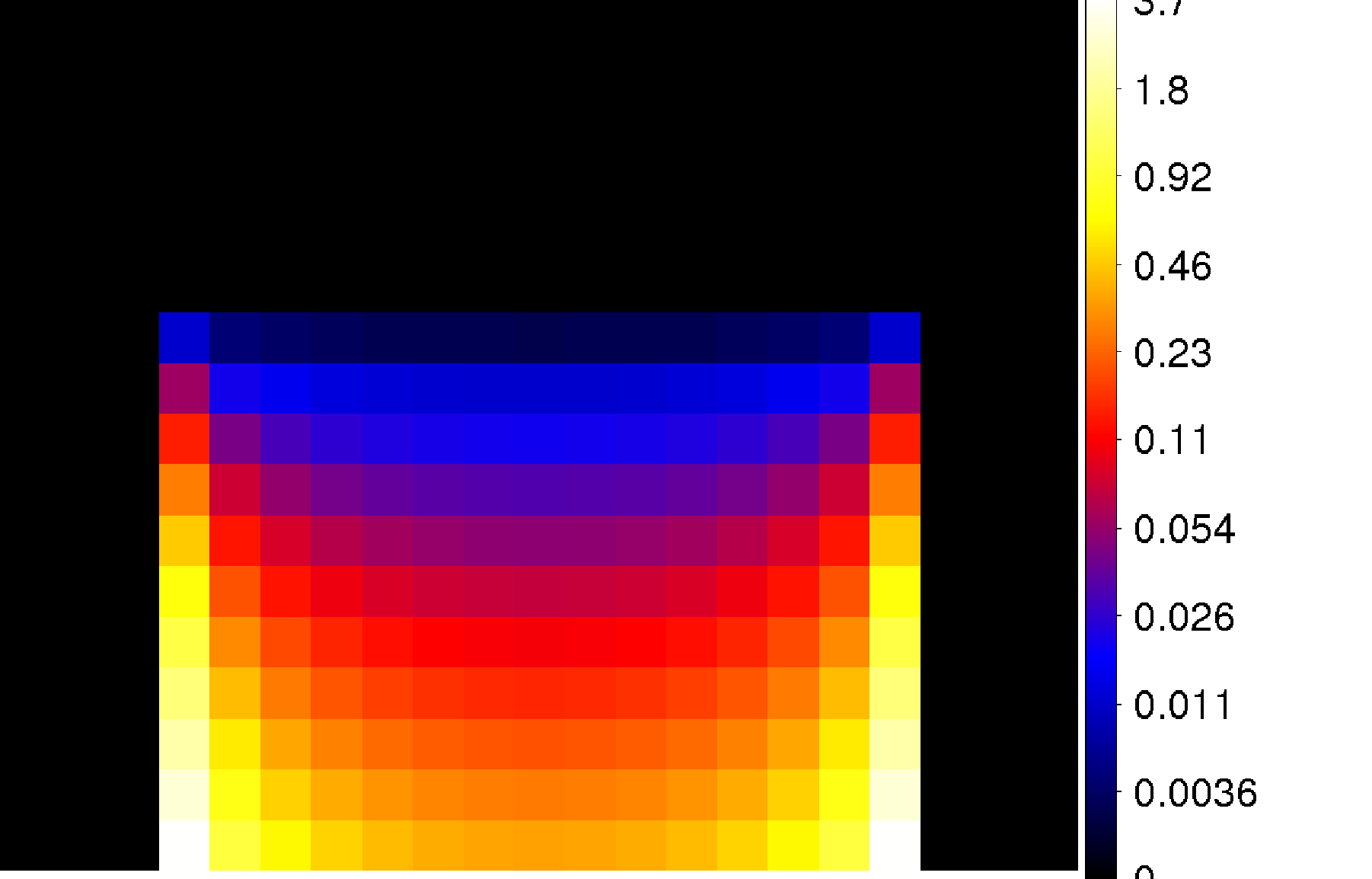 ,width=4.6cm,height=3.2cm}
   \hspace{+0.5truecm}   
                  \epsfig{figure=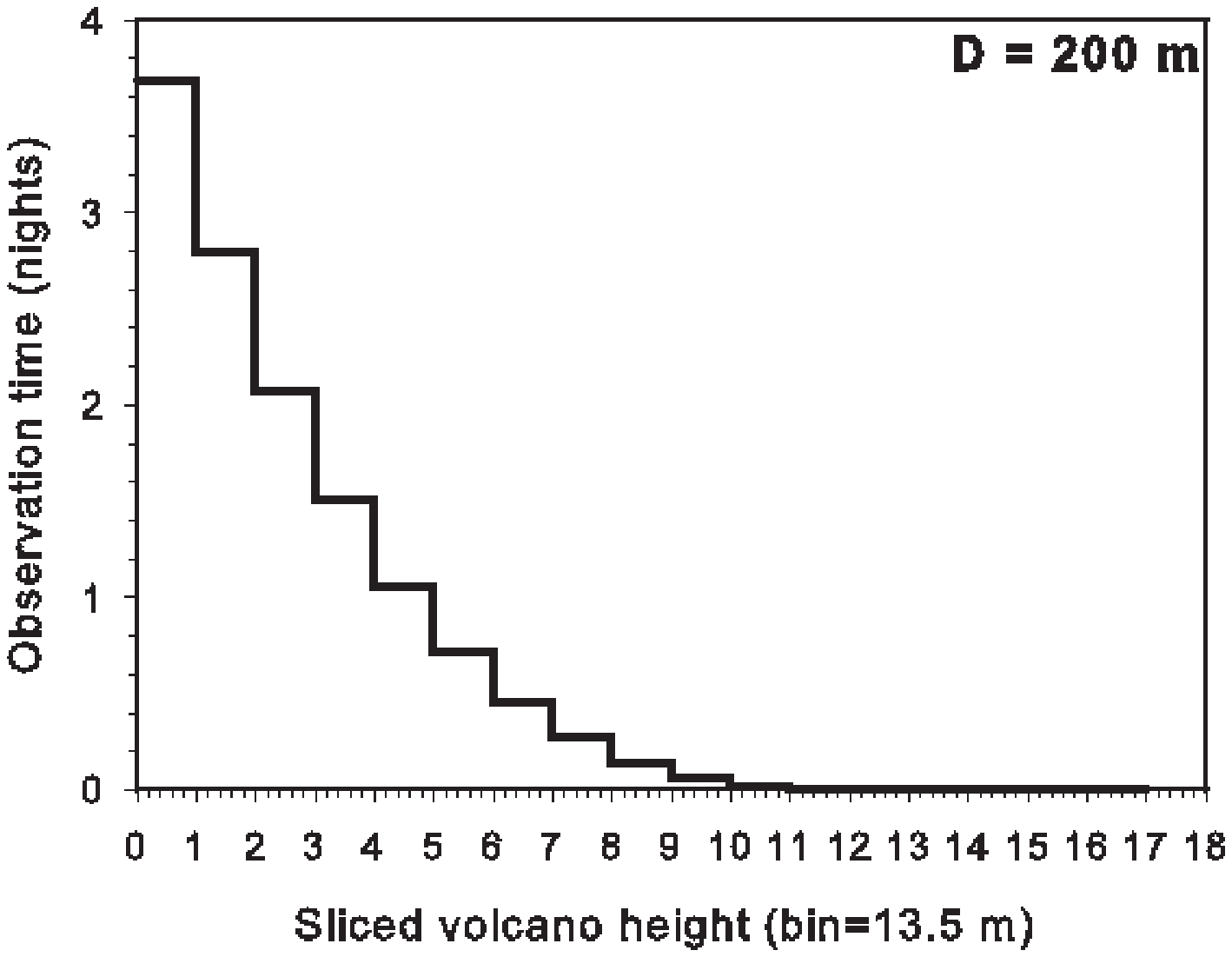,width=4.4cm,height=3.6cm} 
}
\centerline{
  \hspace{-1.2truecm}
                    \epsfig{figure=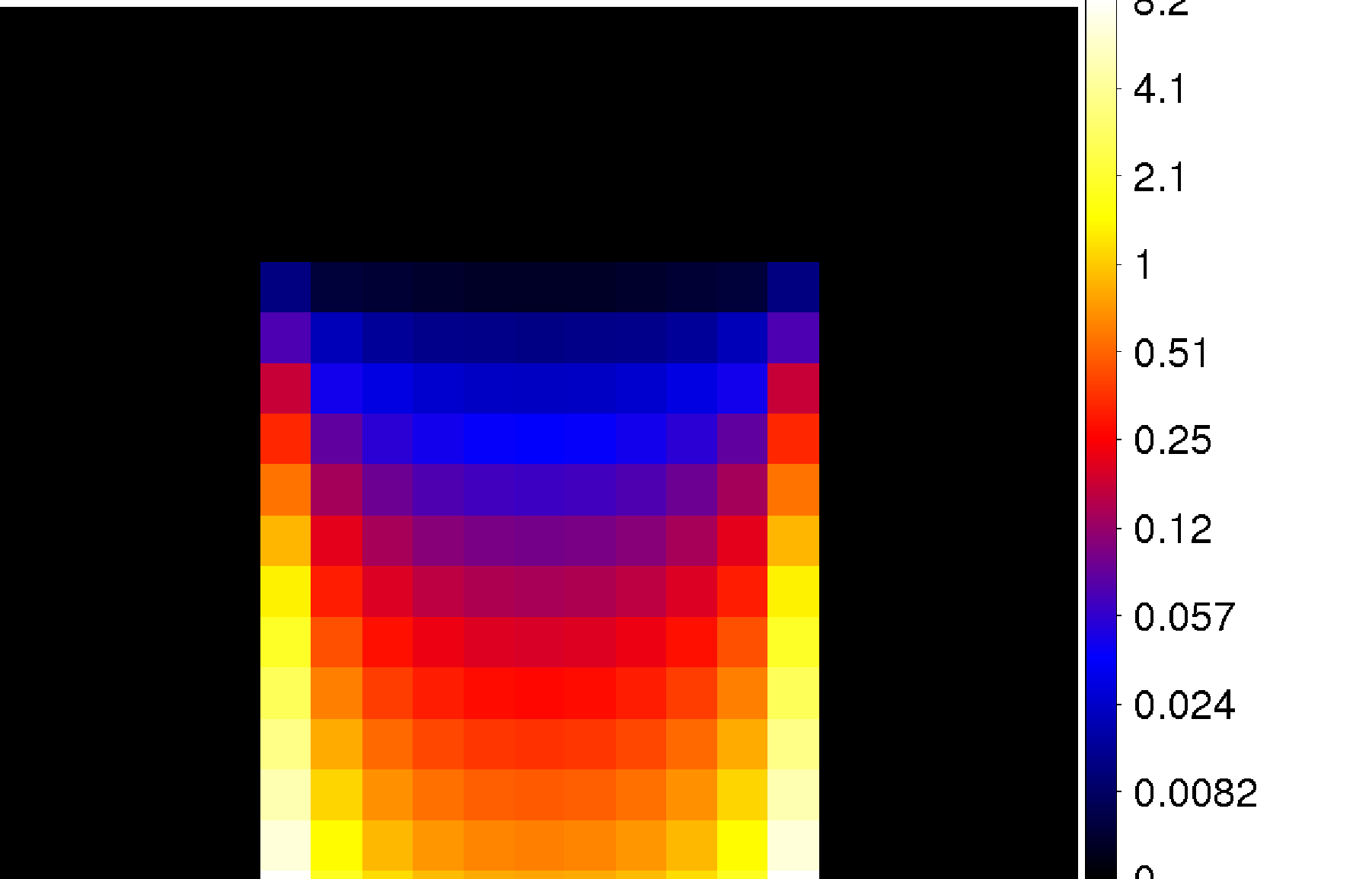,width=4.6cm,height=3.2cm}
    \hspace{+0.5truecm}  
                  \epsfig{figure=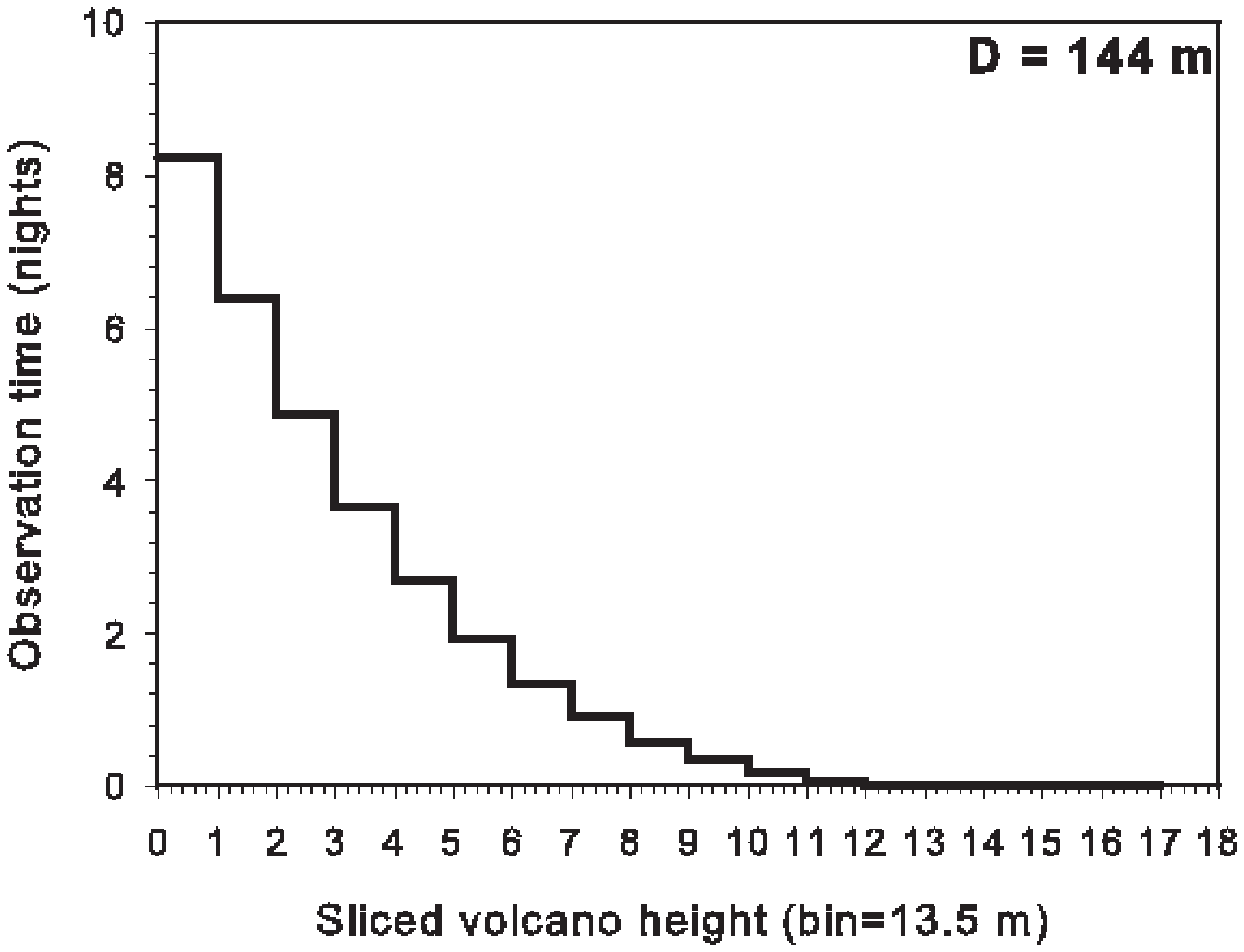,width=4.4cm,height=3.6cm}
 }
\centerline{
  \hspace{-1.2truecm}
                    \epsfig{figure=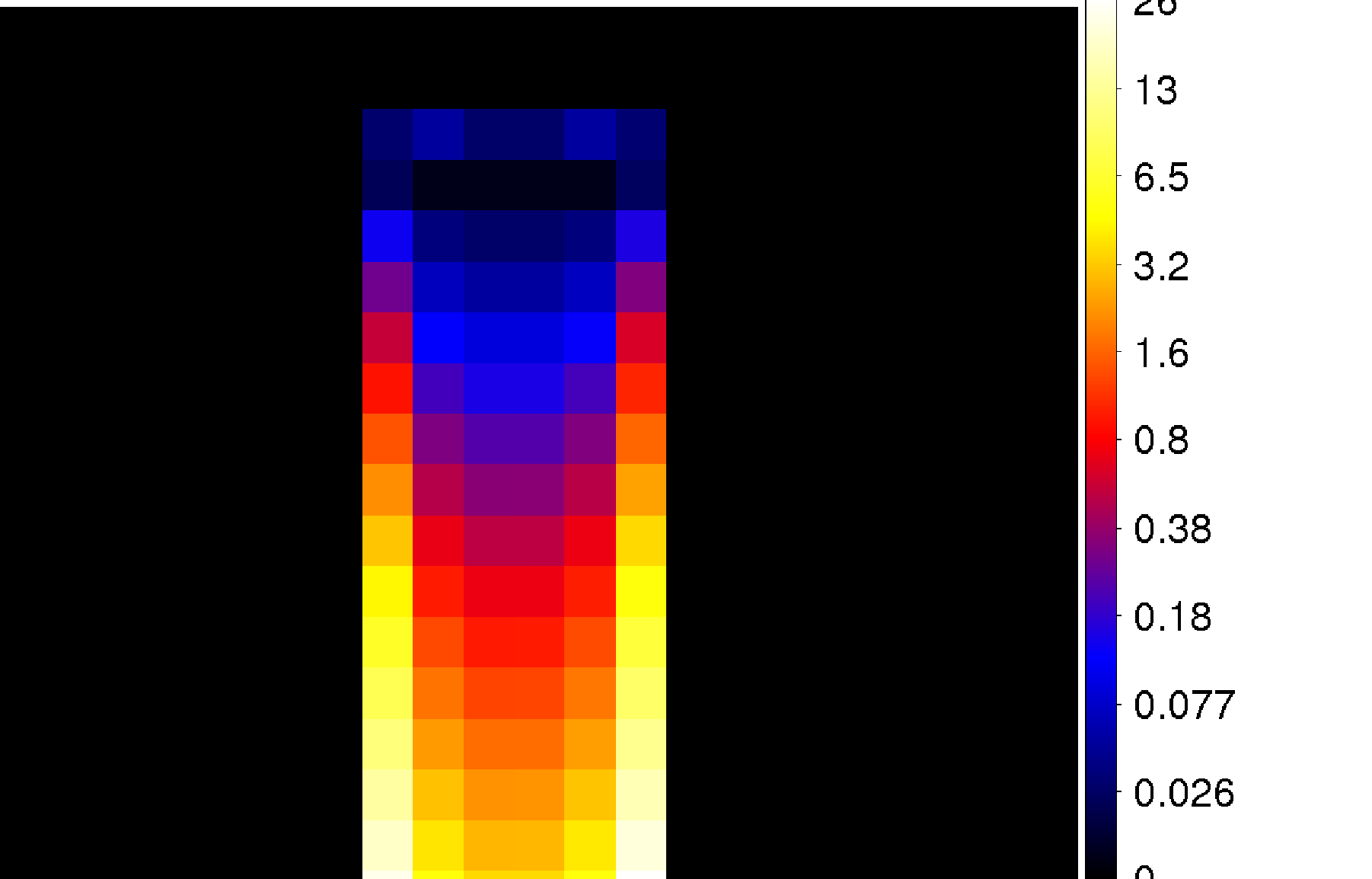,width=4.6cm,height=3.2cm}
   \hspace{+0.5truecm}  
                  \epsfig{figure=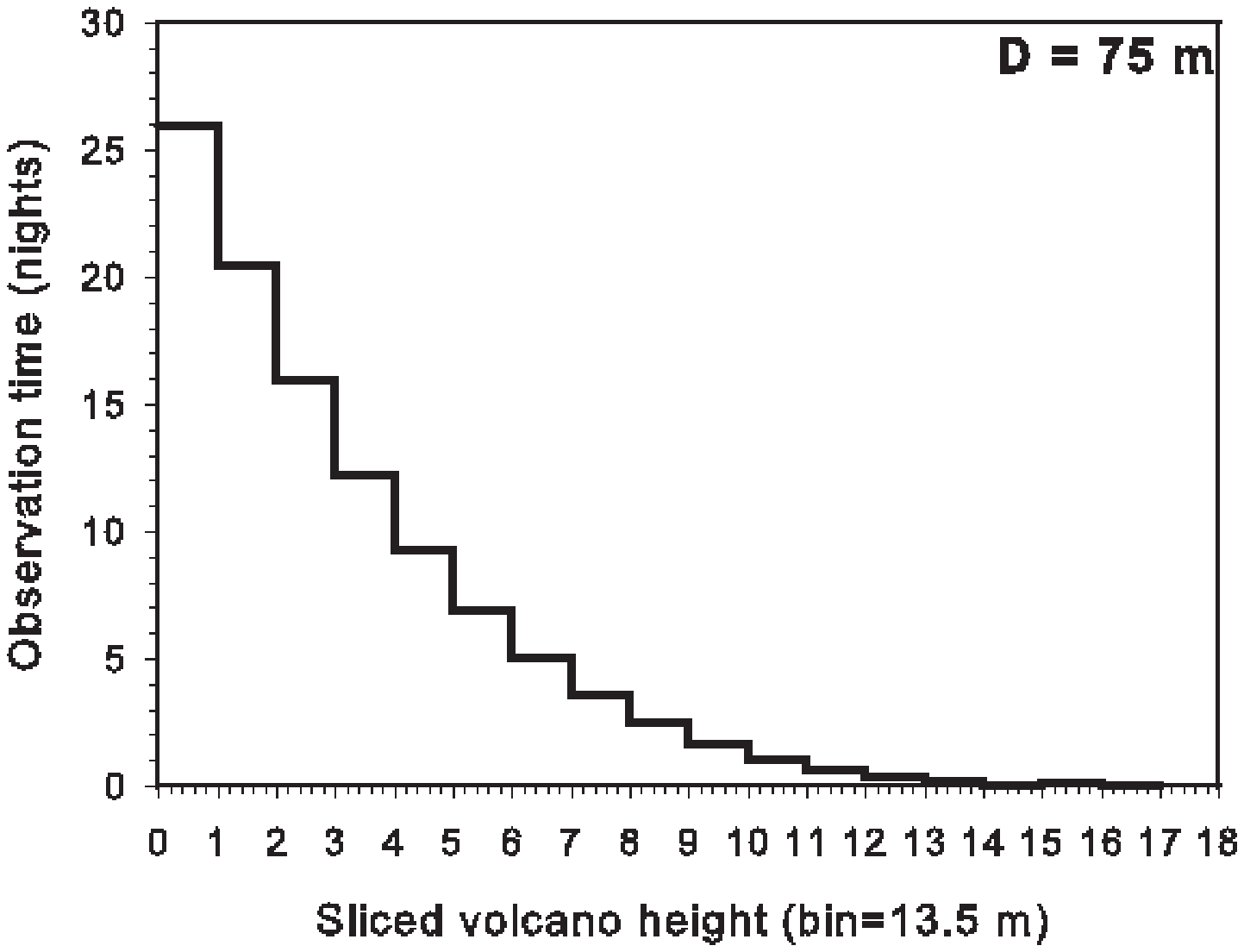,width=4.4cm,height=3.6cm}
 }
\centerline{  
  \hspace{-1.2truecm}
                   \epsfig{figure=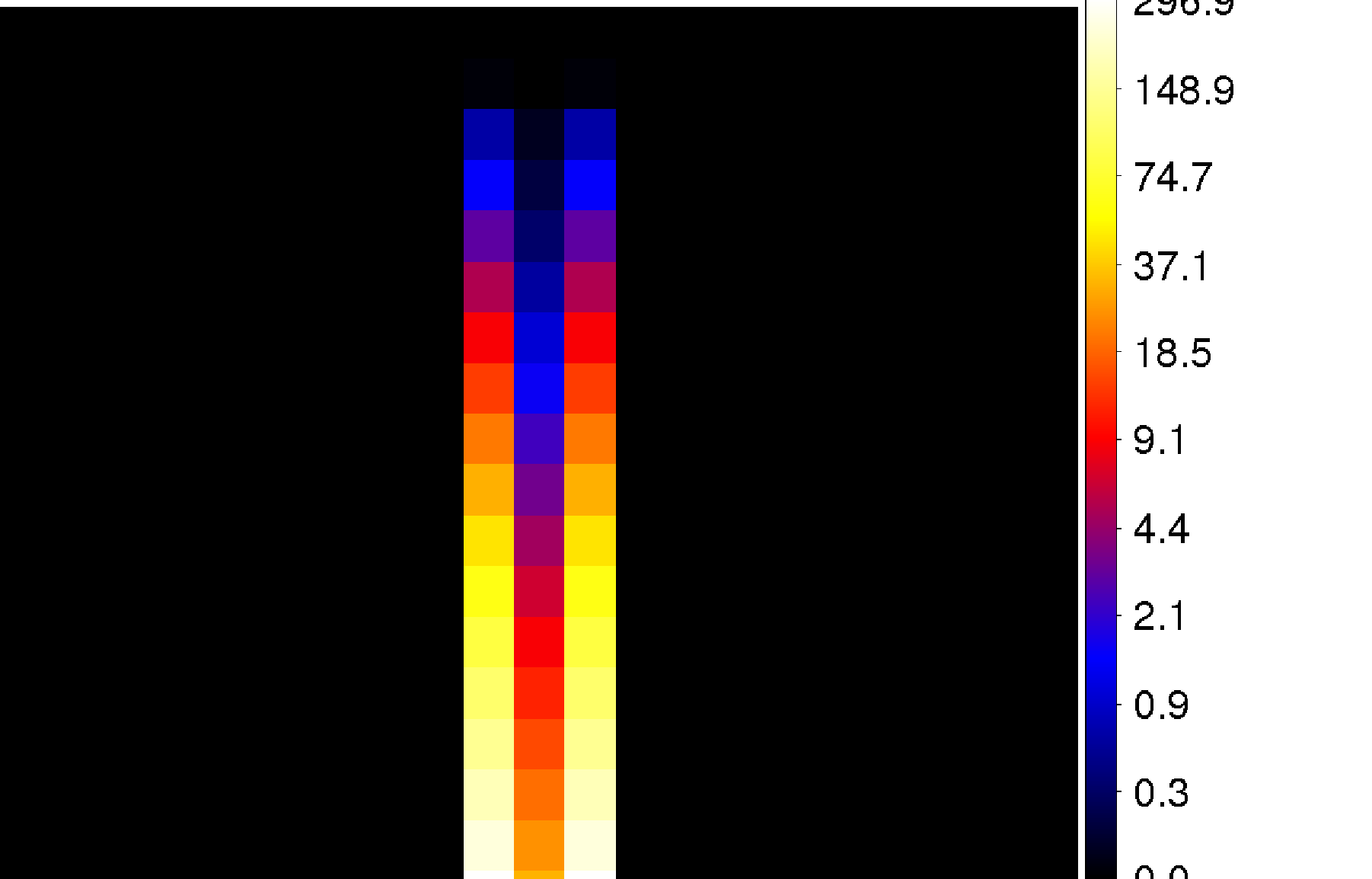 ,width=4.6cm,height=3.2cm}
   \hspace{+0.5truecm}  
                   \epsfig{figure=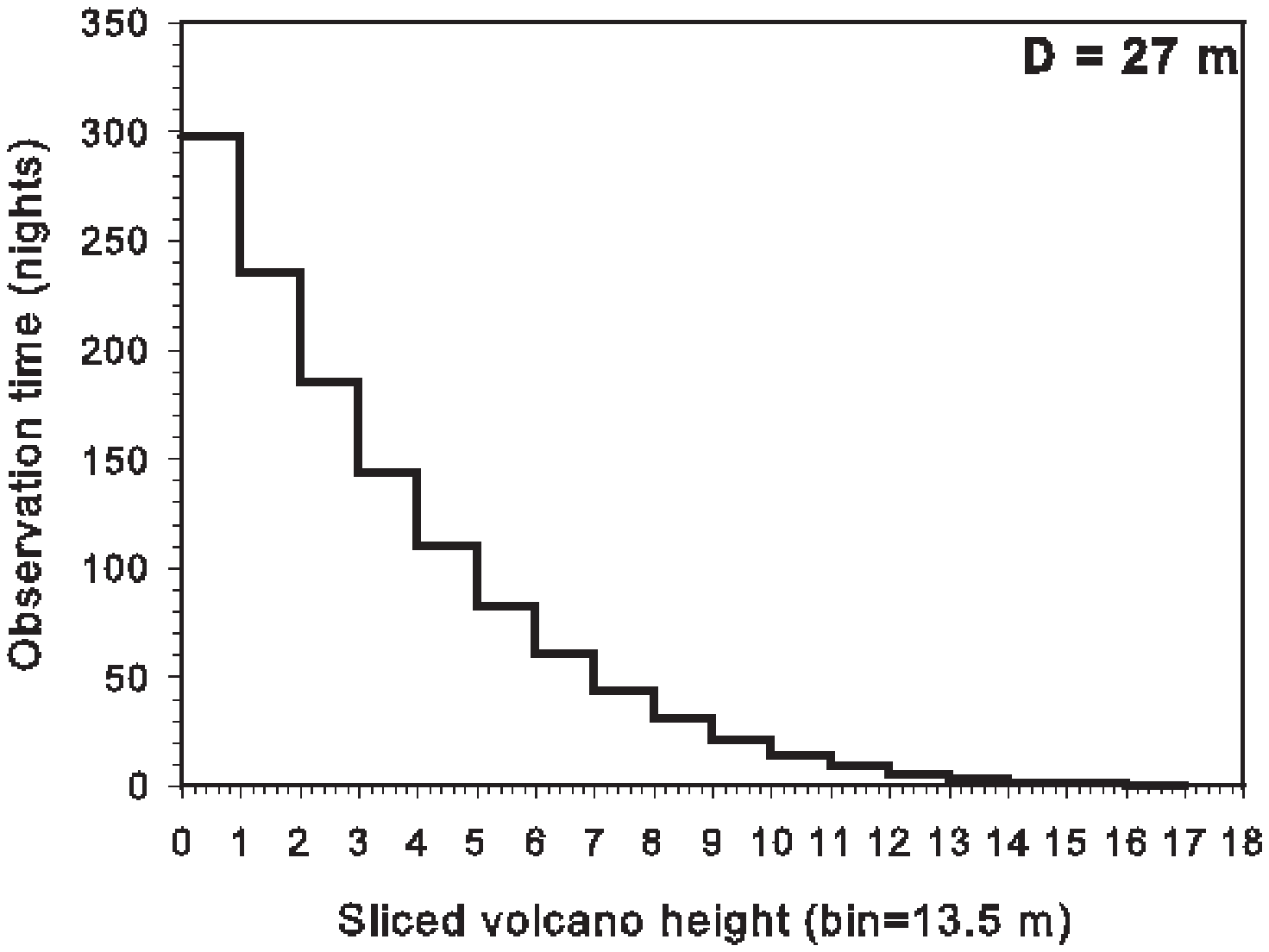 ,width=4.4cm,height=3.6cm}
}
\caption{Minimum observation time  in night units ($\sim$8 hr) necessary to resolve an hollow cylinder inside the volcano. Four different diameters have been considered.
{\it Left}: Spatial distribution in pixels of 13.5$\times$13.5 m$^{2}$; the colours indicates the minimum night values according to the scale 
shown on the right of each figure. {\it Right}: histograms of the observing night as a function of the cone altitude (sliced in 13.5 m bin size).
}
\label{fig:cili_night}
\end{figure}

\begin{figure}[t!]
\centering
\includegraphics[width=12cm]{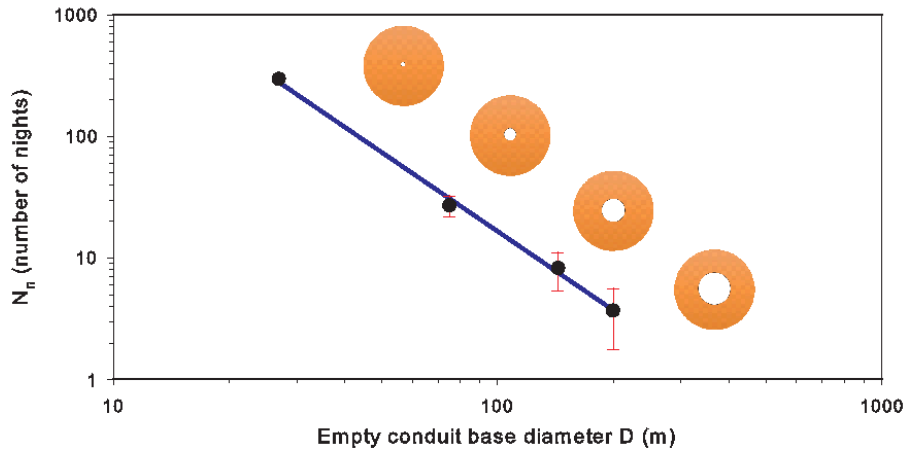}
\caption{Minimum number of observation nights as a function of the conduit diameters. The diameter of each disc represents the cone base, while
the hollow part of each disc is the dimension of the simulated conduit.}
\label{fig:conduit}
\end{figure}

 Assuming as mean duration of 1 night $\sim$8 hr,  the minimum number of observation nights necessary to resolve
 the difference in opacity (at a confidence level  of 68\%) 
 for cylinders of various dimensions (200, 144, 75, and 27 m diameter) is shown in Figures \ref{fig:cili_night}.
Figure  \ref{fig:conduit} shows the minimum number of observation nights (N$_{\rm n}$) as a function of the conduit diameter.
The best fit model of the data is a power-law N$_{\rm n}$=k D$^{-\gamma}$ where k$\approx 3.4\times 10^{5}$  and $\gamma \approx 2.16$.
It is clear that increasing the number of observation nights will allow us to evaluate the difference in opacity with a higher confidence level
and to map the interior of the volcano within the limit of the telescope resolution. 

Similarly, an additional telescope installed in the same site will split in half the observing time necessary to reach the results presented above
and will increase  the sensitivity by a factor of $\sqrt 2$.

The capability to detect the magma filling the conduit is limited by the minimum number of observation nights.
As an example, for a 200 m diameter conduit having a height of H=135 m, 
we obtain from our simulations that a minimum of 3.7 nights (Fig. \ref{fig:cili_night})  is  necessary to resolve the entire conduit.
This value corresponds to a limit average velocity of the magma as $<$v$>$ = $\Delta$H/$\Delta T$ =135 m /3.7 nights $\simeq$ 5 m/h. 
For comparison, the rising magma velocity measured by \cite{tanaka14} for the Satsuma-Iwojima volcano is about 10-30 m/day.

\begin{figure}[h!]
\centering
\includegraphics[width=12cm, height=7.5cm]{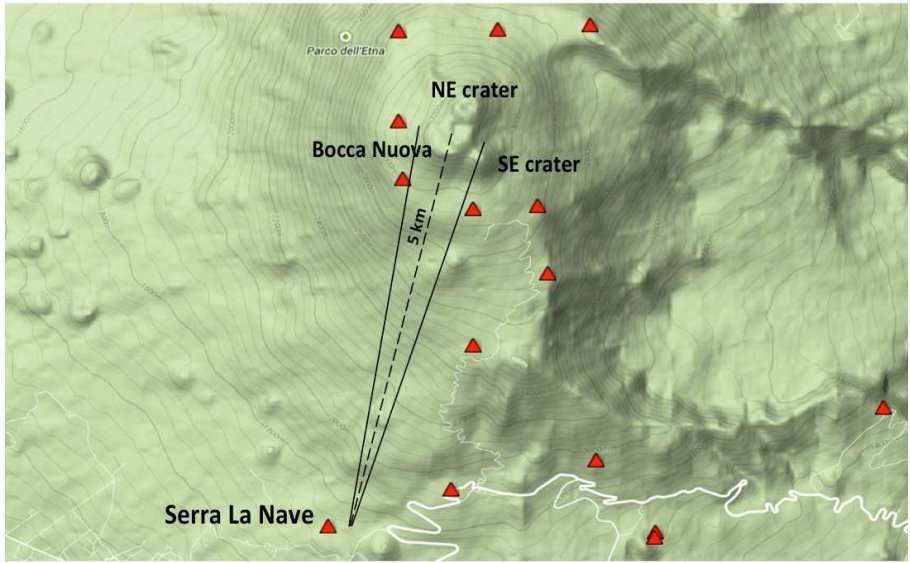}
\includegraphics[width=12cm, height=7.5cm]{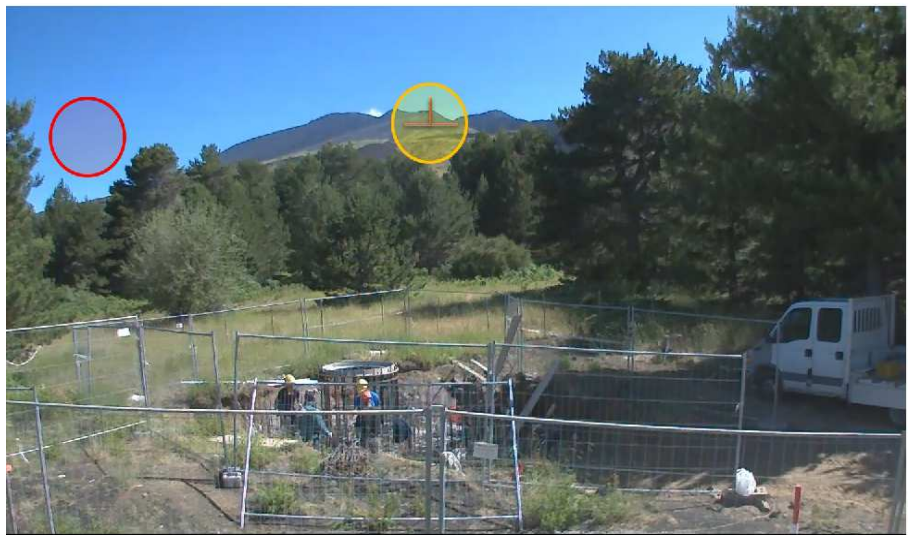}
\caption{{\it{Top}}: Sketch map of the Etna and its three main craters. \astri\ located in Serra La Nave is about 5 km far from the craters. 
{\it{Bottom}}: The yellow circles represent the FOV of the \astri\ telescope imaging the Etna Volcano. Muons crossing the volcano can be detected by 
the \astri\ telescope (the picture shows only the telescope foundation). The red circle represents the FOV directed to open sky for muon flux measurements. }
\label{fig:etna}
\end{figure}

Finally, we have compared the sensitivity of our instrument with the Tanaka experiment.
The sensitivity is given by:
\begin{equation}
S/N =\frac{ {\Gamma \, \Phi_{S} \, T_{\rm obs}} }{\sqrt{\Gamma \, \Phi_{B} \, T_{\rm obs}}}
\label{eq:sensitivity}
\end{equation}
\noindent
where  $\Phi_{S}$ and $\Phi_{B}$ are the source and background flux, respectively, and $T_{\rm obs}$ is the observing time.
Assuming the background an acceptance reported in \cite{tanaka14}  ($\sim10^{-3}$ cm$^{-2}$ sr$^{-1}$ day$^{-1}$ and $\approx$3650 cm$^{2}$ sr, respectively)
and the \astri\ values reported in this paper ($\sim 3 \times10^{-6}$ cm$^{-2}$ sr$^{-1}$ day$^{-1}$ and $\approx$2860 cm$^{2}$ sr),
we roughly estimate that we gain a factor of ten in sensitivity  compared with the experiment performed by  \cite{tanaka14}.
It is worth noting that the sensitivity comparison between the two experiments  takes also into account that the Tanaka particle detector can observe even during the day.

\section{Validation of the method using \astri}
When operative, \astri\ will be able to test  the new method we have proposed for volcano muon radiography.
The \astri\ telescope can be pointed towards the volcano peak located at distance of about 5 km, 
catching into a single view the entire main crater area as shown in Figure \ref{fig:etna} (top).
The Etna volcano has a base diameter of about 40 km and a height of about 3350 m. 
One of the active craters (SE Crater; see Fig. \ref{fig:etna}, bottom) in the summit area of the volcano will be the target of our experiment of muon radiography. 
Muon particles crossing the Etna's SE crater and reaching \astri\ would have a zenith angle of about 76\deg\ and would cross up to 500 m of rock.

The \astri\ location allows us to perform the radiography of the volcano with a spatial resolution of 
about 15 m. Considering the total \astri\ acceptance $\Gamma = 2860$ cm$^{2}$ sr, where the collecting area of the telescope  is $\approx$13 m$^{2}$ 
and the half FOV is 4.8\deg, a total of $\approx$5000 muons/night, including those arriving from open sky, are expected to hit the telescope mirror. 

A preliminary measurement of muon flux will be carried out
pointing the telescope to the open sky at equal
elevation angle as the volcano (Fig. \ref{fig:etna}), bottom) and compared with the flux measured
pointing  to the volcano. The ratio between the two fluxes  gives the estimate of muon
attenuation inside the volcanic structure.

\section{From muon radiography to muon tomography}
\begin{figure*}[t!]
\begin{center}
\includegraphics[width=10cm,height=7cm]{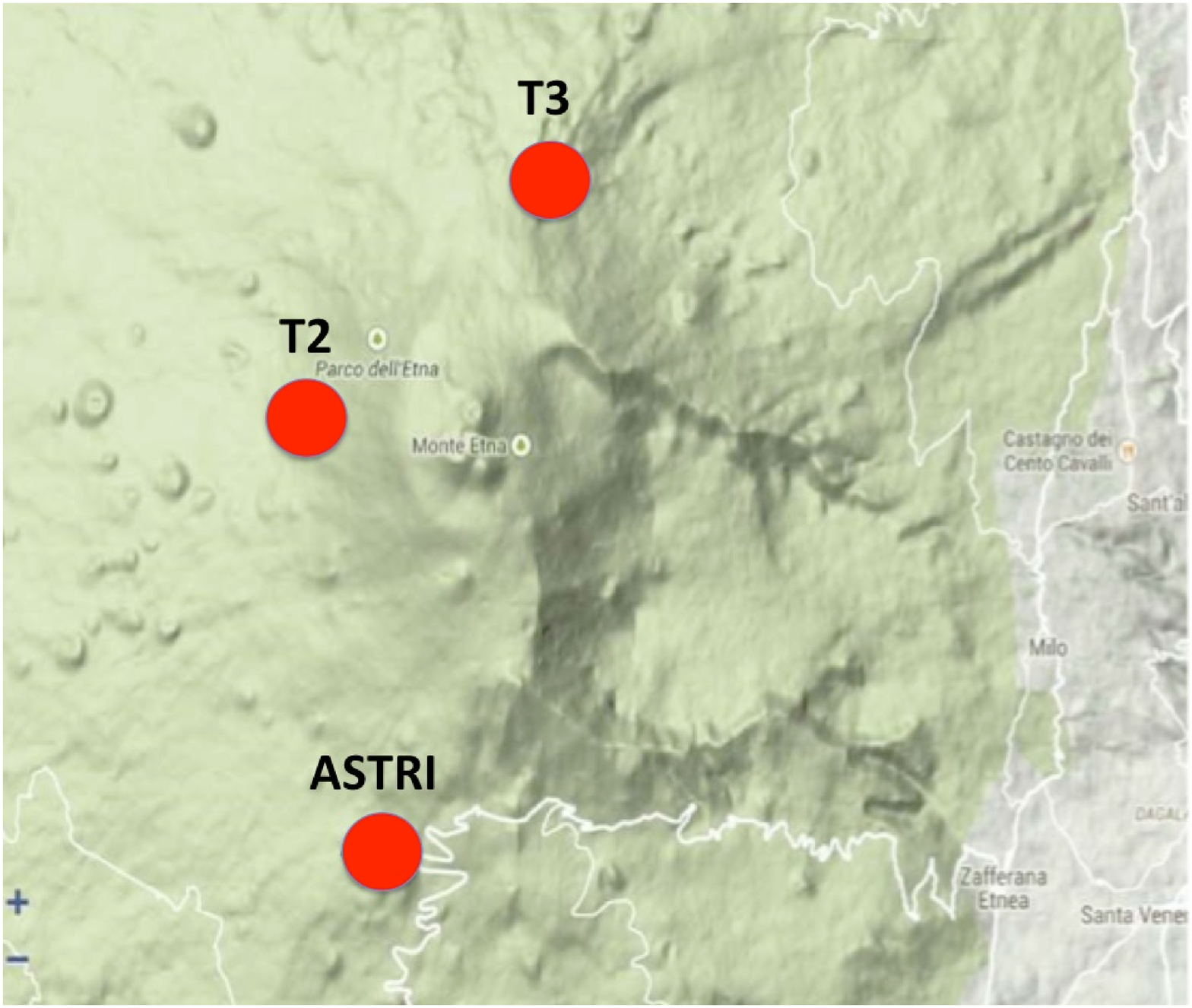}
\includegraphics[width=10cm, height=7cm]{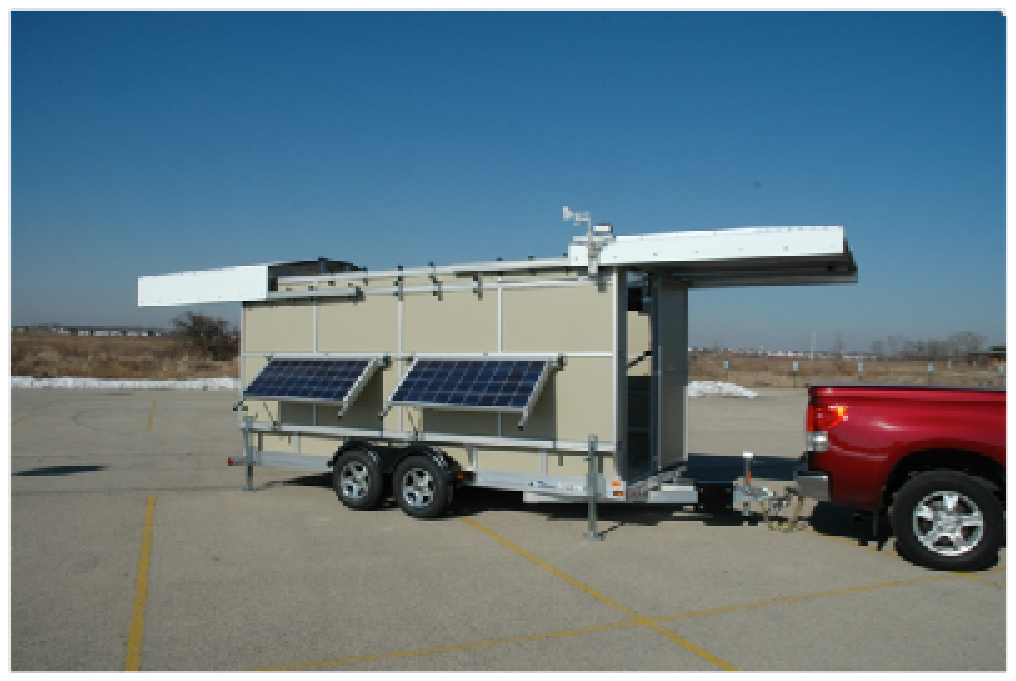}
\end{center}
\caption{
{{\it {Top}}:  A possible Cherenkov telescopes configuration for the Mt. Etna tomography.
{\it {Bottom}}: An example of trailer that could be used for hosting the telescope and moving it in different Etna observing sites. }
}
\label{fig:tomo}
\end{figure*}

Our simulations have shown that muon radiography using Cherenkov technique will allow us to measure, 
with higher resolving power and negligible background, the internal density structure of volcanoes. 
Measuring the muon flux absorption as a function of the muon direction 
resolves only the average density distribution along individual muon paths. 
Obviously, several telescopes at different position around the volcano would be required
to distinguish between an empty cavity or a relatively low density region.
Multidirectional radiography (tomography) can resolve the exact position of the density anomaly,
 its shape and its alignment by superimposing images obtained by each telescope and producing
 three dimensional images of the region of interest. 
 For muon tomography, the \astri\ telescope should be complemented with at least two other Cherenkov telescopes.
 Such telescopes do not need to be ASTRI-like and do not need either a pointing system or a double mirror.
 They will be dedicated only to the detection of muon rings and designed for this purpose. 
The additional two telescopes may be installed on mobile vehicles and solar powered, 
 so that they can be positioned in desired areas and operated remotely (Fig. \ref{fig:tomo}, bottom). 
 The three telescopes have to be located around the Etna volcano on the vertex of a triangle containing the volcano (Fig. \ref{fig:tomo}, top).  
 Such an array of telescopes exploiting the $\mu$-Cherenkov technique could make the difference in the field 
 of volcano tomography and open a new frontier in volcanology.

\section{Conclusions}
 Mapping the interior of active volcanoes 
could help to interpret warning signs and improve prevention for volcanic events.
Far to be exhaustive in all its part, this paper covers the main feasibility aspects of the 
new proposed method for Cherenkov muon imaging applied to volcanology. 
Our toy-model simulations show that we gain at least a factor of ten in sensitivity when comparing
 with previous experiments.

We will be able to test this method with \astri\ thanks to its favoured position on the Mt Etna.
However, we hope that a configuration with more than one telescope, allowing also for muon tomography, 
can be achieved and supported by the field experts at large.

\section*{Acknowledgements}
This work was partially supported by the ASTRI Flagship Project financed by the Italian Ministry of Education, 
University, and Research (MIUR) and led by the Italian National Institute of Astrophysics (INAF). 
We also acknowledge partial support by the MIUR Bando PRIN 2009 and TeChe.it 2014 Special Grants.
We area grateful to the entire ASTRI collaboration and to the INAF HQ for encouraging the present work. 
We thanks the colleagues of the Catania Astrophysical Observatory-INAF and of the Etna Regional Park for many useful discussions.

\section*{References}
  \bibliographystyle{elsarticle-num} 

\begin{thebibliography}{00}
\bibitem{nagamine03} K. Nagamine, 2003, Introductory Muon Science, Cambridge University Press, Cambridge, 208pp
\bibitem{george55}  E. P. George, 1955, Commonwealth Engineer, 455, 457.
\bibitem{alvarez70}  L.W.  Alvarez, et al., Science, 167 (1970) 832
\bibitem{larocca15} P. La Rocca et al., Nucl. Instrum. and Meth. A, 787, 236
\bibitem{heck98} D. Heck, et al., 1998, Forschungszentrum Karlsruhe Report FZKA, 6019
\bibitem{bugaev98} E. V. Bugaev, et al., 1998, Phys. Rev., 58, 05400 
\bibitem{darijani14} R. Darijani, et al. , 2014, Ind. Journ. of Pure \& Applied Physics, 52, 7 
\bibitem{lesparre10} N. Lesparre, et al., 2010, Geophys. J. Int., 183, 1348
\bibitem{voelk09} H. J. Voelk \& K. Bernloehr, 2009, Exper. Astron.,  25, 173
\bibitem{tanaka07}  H.K.M. Tanaka, et al., 2007, Earth planet. Sci. Lett., 263,104
\bibitem{tanaka09}  H.K.M. Tanaka,  et al.,  2009, Geophys. Res. Lett., 36, L01304
\bibitem{nagamine95} Nagamine, K., et al., 1995, Nucl. Instrum. Methods, 356, 585
\bibitem{tanaka01}  H.K.M. Tanaka,  et al.,   2001, Hyperfine Interact., 138, 521
\bibitem{tanaka03} H.K.M. Tanaka,  et al.,  2003, Nucl. Instr. and Meth. A, 507, 657
\bibitem{tanaka14} H.K.M. Tanaka, et al., Nature Comm. 5 (2014) 3381
\bibitem{ambrosi11} G. Ambrosi, et al., 2011, Nucl. Instr. and Meth. A, 628, 120
\bibitem{anastasio13} A. Anastasio, et al., 2013, Nucl. Instr. and Meth. A, 718, 134
\bibitem{carbone13} D. Carbone, et al., 2013, Geophys. Jou. Int, 58, 054001
\bibitem{weekes02} T.C. Weekes, et al., 2002,  Astroparticle Physics, 17, 221
\bibitem{tridon10} Tridon, D. B., et al., 2010, NIMPA, 623, 437
\bibitem{benbow05} W. Benbow, AIP Conference Proceedings, 745, 2005, 611
\bibitem{actis11} M. Actis,  et al., 2011, Exp. Astronomy 32, 193 
\bibitem{vacanti94} G. Vacanti, et al.,1994, APh, 2, 1
\bibitem{rovero96} Rovero, A. C., et al., 1996,  Astrop. Phys., 5, 27
\bibitem{meyer05} Meyer, M., et al., 2005,  American Institute of Physics Conference Series, 745, 774
\bibitem{bolz04} Bolz, O., 2004,  Ph.D. thesis, Karl-Ruprecht University, Heidelberg, Germany
\bibitem{vercellone13} S. Vercellone, et al., 2013, Fermi Symposium proceedings, arXiv:1303.2024
\bibitem{catalano14} O. Catalano, et al., 2014, Proc. of the SPIE, 9147, 91470
\bibitem{canestrari14}  R. Canestrari, et al., 2014, Proc. of the SPIE, 9145, 91450
\bibitem{catalano13}  O. Catalano, et al., 2013, Proc. 33rd ICRC2013, arXiv:1307.5142
\bibitem{maccarone13}  M. C. Maccarone, et al., 2013, Proc. 33rd ICRC, arXiv:1307.5139
\bibitem{leto14}  G. Leto, et al., 2014, Proc. AtmoHEAD Workshop 2013, arXiv:1402.3515v1
\bibitem{strazzeri13}  E. Strazzeri, et al., 2013, Proc. of the 33rd ICRC, arXiv:1307.5204
\bibitem{grieder01} P. K. F. Grieder, 2001, Cosmic Rays at Earth: Researcher's Reference Manual and Data Book, Eds. Elsevier


\end{thebibliography}


\end{document}